\documentclass[twocolumn]{aastex631}

\begin{document}

\title{Lithium as a Signpost for Compact Object Binary Candidates in the LAMOST Medium Resolution Survey}

\author[0009-0006-7701-9174]{Yang Shen}
\affiliation{Harvey Mudd College, 301 Platt Blvd, Claremont, CA 91711, USA}
\altaffiliation{Corresponding author: \href{mailto:lshen@hmc.edu}{lshen@hmc.edu}}

\author[0000-0002-4733-4994]{Joshua D. Simon}
\affiliation{Observatories of the Carnegie Institution for Science, 813 Santa Barbara St., Pasadena, CA 91101, USA}

\author[0000-0002-6871-1752]{Kareem El-Badry}
\affiliation{Department of Astronomy, California Institute of Technology, 1200 E. California Blvd., Pasadena, CA 91125, USA}

\author[0000-0002-6406-1924]{Casey Y. Lam}
\affiliation{Observatories of the Carnegie Institution for Science, 813 Santa Barbara St., Pasadena, CA 91101, USA}



\begin{abstract}
Binary systems with black hole or neutron star companions are often associated with lithium enhancement. Gaia NS1, a recently discovered neutron-star binary with a lithium-enhanced main-sequence companion, demonstrates the potential of lithium as a signpost for identifying compact object binaries in existing spectroscopic surveys. In particular, we aim to use lithium as a signpost to find compact object binaries similar to Gaia NS1 in the Large Sky Area Multi-Object Fibre Spectroscopic Telescope (LAMOST) Medium Resolution Survey (MRS). From LAMOST MRS, we selected 4441 metal-poor main-sequence stars like Gaia NS1, measured the Li 6707 \r{A}  equivalent width, and then identified a sample of 33 stars with strong Li absorption. We used radial velocity variation and astrometric binarity signatures from Gaia, narrowing the sample to 3 candidates. We identified one of these candidates as an eclipsing binary and demonstrated that massive companions for the other two are unlikely via follow-up spectroscopy. 

\end{abstract}

\section{Introduction}\label{intro}
Binary systems with black hole (BH) or neutron star (NS) companions are fundamental for understanding the outcome of supernova explosions and the production of these compact objects from the cores of massive stars. Searches for binaries with invisible companions generally rely on radial velocity (RV) or astrometric monitoring. \citep[e.g.,][]{2014Natur.505..378C, 2022A&A...658A.129J}. Lithium enhancement may provide a novel way to identify compact object binaries.

Li enhancement is commonly seen in the companion star for BH and NS binary systems \citep[e.g.,][]{2007A&A...470.1033C, 2015NewAR..64....1L}. After Gaia Data Release 3 \citep[DR3;][]{2023A&A...674A...1G}, \citet{2024OJAp....7E..27E} confirmed that Gaia NS1 hosts a NS and found that its main-sequence companion has a significantly higher Li abundance than stars with similar stellar parameters. Motivated by Gaia NS1, we searched spectroscopic surveys for similar compact object binaries.

\section{Data Analysis} \label{data analysis}
\subsection{Primary sample selection}
Our sample was obtained from the Large Sky Area Multi-Object Fibre Spectroscopic Telescope (LAMOST) Medium Resolution Survey (MRS) Data Release 9 \citep{2012RAA....12.1197C}. We selected a primary sample of 4441 Gaia NS1-like stars with $\log{g} > 4$, $[\mathrm{Fe/H} ] < -1$, a surface temperature between 4000 and 7000~K, and $\mathrm{S/N > 10}$. 

\subsection{Lithium abundance estimation and targets selection}
We used the H$\alpha$ absorption line to estimate the RV for each LAMOST spectrum, from which we determined the expected location of the \ion{Li}{1} 6707~\AA\ absorption line. Then, we performed a Gaussian fit, fixed the Gaussian width to the LAMOST spectral resolution ($R \approx 7500$), and constrained the amplitude to be negative. We derived the Li equivalent width (EW) and its uncertainty ($\sigma_{\mathrm{EW}}$) from the Gaussian fit.

To select Li-enhanced stars, we first took the weighted average of the Li EW and uncertainty for all spectra of each star. We fit a Gaussian function to the EW distribution. For $|\text{EW}| = 0.13$ \r{A}, the EW distribution significantly exceeds the Gaussian fit. We thus adopt $\mathrm{|EW| > 0.13}$\r{A} as a conservative threshold for Li enhancement. We additionally required $\mathrm{EW}/\sigma_{\mathrm{EW}}>3$ to ensure accuracy, yielding a final sample of 33 lithium-enhanced stars. 

\subsection{Tests for binarity}
Next, we search for evidence of binarity for these stars using two approaches: (1) the $\chi^2$ test for RV variability and (2) the Renormalized Unit Weight Error (RUWE) from Gaia.

Binary stars exhibit RV variability. We can measure the significance of variation using the $p$-value from a $\chi^2$ test of the RVs. We conduct this test with the RV measurements from LAMOST MRS and Gaia DR3. We set $p < 0.05$ as a threshold for a star exhibiting variable RVs. The typical RV uncertainties for the 33 Li-enriched stars are $2-4$ km~s$^{-1}$, so this test is sensitive to binaries with RV variations $>5-10$ km~s$^{-1}$, and companions inducing smaller amplitudes may remain undetected.

The RUWE quantifies how static the star's position is compared to the single-source astrometric solution. Therefore, high RUWE may indicate the presence of an unseen companion. We took the average of the suggested RUWE thresholds  from \cite{2022MNRAS.513.2437P} and \cite{2024A&A...688A...1C} and set our threshold for a high likelihood of binarity as 1.20.

Based on the $\chi^2$ test for RVs and RUWE threshold, 3 out of the 33 Li-enhanced stars show evidence for binarity: J000045.08+062941.8 (Gaia DR3 2745419790836631424), J000556.86$-$012835.5 (Gaia DR3 2448812029433640704), and J065401.91+752725.9 (Gaia DR3 1115628342232842368). 

\subsection{Light curve data}
In addition to RVs and astrometric parameters for assessing binarity, we obtained light-curve data from the Transiting Exoplanets Survey Satellite \citep{2015JATIS...1a4003R} for our 3 candidates. We identified candidate periods with Lomb–Scargle periodograms and phase-folded the light curves. The light curve of one star, J000045.08+062941.8 (TIC 418735227), exhibits eclipses with a period of 2.21 days. We classify it as an eclipsing binary containing 2 luminous stars, ruling out a compact-object companion. However, we note that the RUWE value of 1.31 for this star could suggest the presence of a wider tertiary companion. The other two candidates show no significant photometric variability.

\section{Spectroscopic Follow-up}\label{additional_RV}
To assess the remaining candidates' nature, we obtained 5 additional spectra of J065401.91+752725.9 with the Automated Planet Finder (APF) at Lick Observatory \citep{2014PASP..126..359V} and 5 spectra of J000556.86-012835.5 with the MIKE spectrograph on the Magellan Clay telescope \citep{2003SPIE.4841.1694B} and APF. We measured RVs from these spectra using the same methods as in \cite{2026arXiv260320371S}. 

\begin{figure*}[!h]
\centering
    \includegraphics[width=\textwidth]{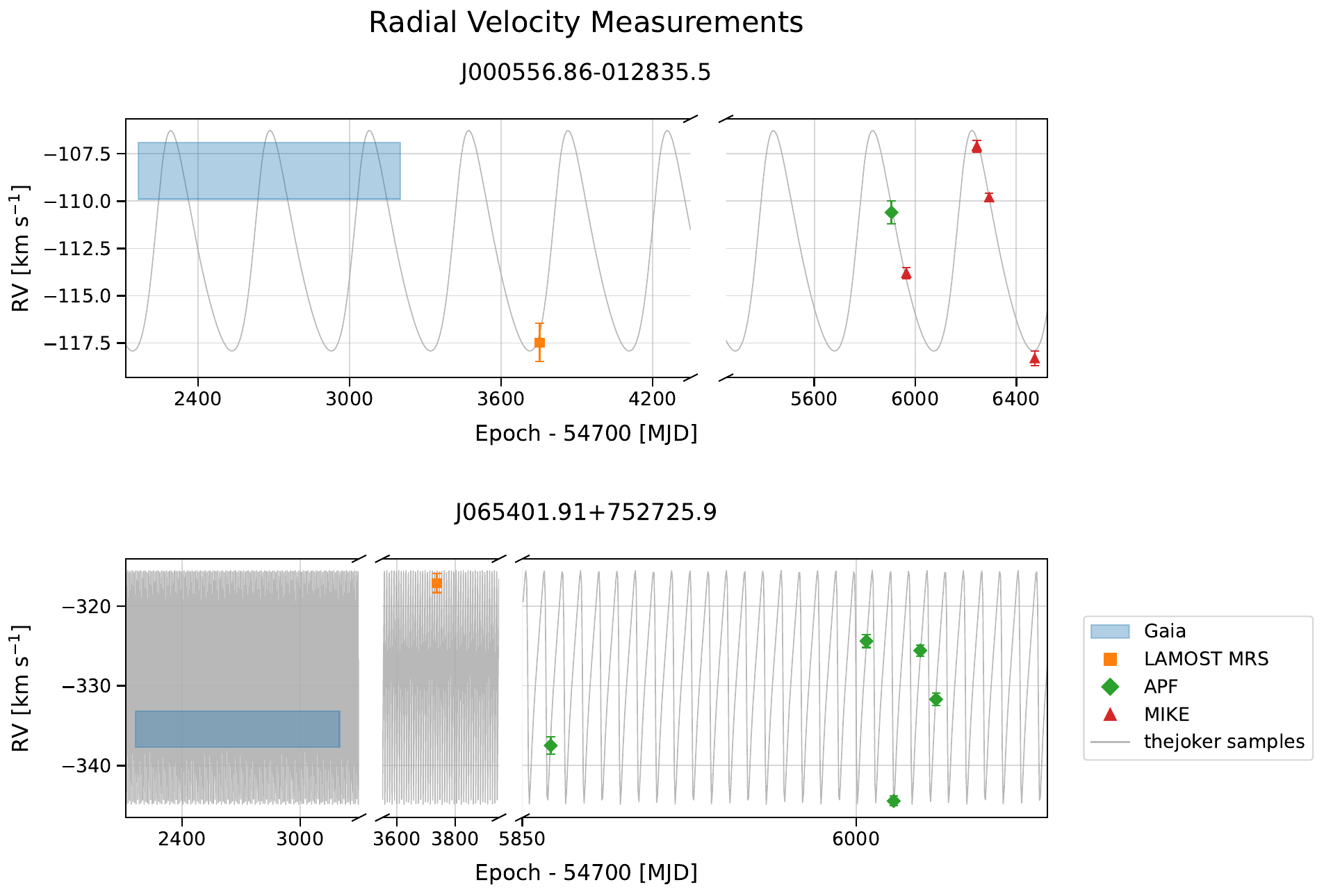}
    \caption{The radial velocity measurements for the lithium-enriched binary candidates from Gaia, LAMOST MRS, APF, and MIKE. Gaia RVs are shown as a band to represent the 34-month averaging interval. Example \texttt{thejoker} orbital solutions are overplotted.. The example solution for J000556.86-012835.5 has $P = 393.34$ days and $K = 5.82$ km~s$^{-1}$. The one for J065401.91+752725.9 has  $P = 8.19$ days and $K = 14.66$ km~s$^{-1}$. }
    \label{fig:radial velocity}
\end{figure*}

We combined these spectra with LAMOST RVs and fit orbital solutions using \texttt{thejoker} \citep{2017ApJ...837...20P}. For J000556.86-012835.5, the samples indicate a most likely period of $\mathrm{P} \sim 400$ days and semi-amplitudes $K$ from $\sim 5-10$ km~s$^{-1}$. For J065401.91+752725.9, we find that potential periods span $3-30$~days with $K$ from $12-53$~km~s$^{-1}$. Since J065401.91+752725.9 has a relatively short period compared to the Gaia DR3 duration, we can use the Gaia rv\_amplitude\_robust an approximation for $2K$. Given this limit, we can reject the solutions with $K > \sim 20$ km~s$^{-1}$, which will result in larger Gaia amplitudes than observed.  However, the period $P$ remains unconstrained from $\sim 3-30$ days. Figure \ref{fig:radial velocity} shows the RVs and example \texttt{thejoker} solutions.

Using the binary mass function and assuming the main sequence stars have masses equal $0.8 M_{\odot}$, we found that the inferred companion masses for both systems are too small to correspond to BHs or NSs (J000556.86-012835.5: $\sim0.19 M_{\odot}$, J065401.91+752725.9: $\sim0.11 M_{\odot}$ using the example solutions in Figure \ref{fig:radial velocity}). Compact object companions would require nearly face-on inclinations.

Therefore, although strong lithium absorption motivated our search for NS/BH binary systems, none of our three candidates exhibits evidence of NS/BH companions. The physical origin of the excess lithium found in these systems remains a mystery.

\section{Code release and additional spectroscopic surveys}
The analysis pipeline is released on \href{https://doi.org/10.5281/zenodo.20769778}{Zenodo}. We have also applied the same pipeline to other catalogs, including the LAMOST Low Resolution Survey DR9, the Sloan Digital Sky Survey DR16, the GALactic Archaeology with HERMES DR4, and the Dark Energy Spectroscopic Instrument DR1. However, we did not identify any strong candidates for binaries with high Li abundances in those data sets.

\section*{Acknowledgments}
Guoshoujing Telescope (LAMOST) is a National Major Scientific Project built by the Chinese Academy of Sciences. Funding for the project has been provided by the National Development and Reform Commission. LAMOST is operated and managed by the National Astronomical Observatories, Chinese Academy of Sciences.

\newpage
\bibliography{compact_object}{}
\bibliographystyle{aasjournal}



\end{document}